# The Cardan grille approach to the Voynich MS taken to the next level

*René Zandbergen*

*(Independent researcher)*

## Abstract

The Voynich MS is an illustrated 15[th] century manuscript, whose text is written in an unknown alphabet, which has not been translated until today. In 2004 Gordon Rugg published a paper in which he proposed that this text is likely to be meaningless, and could have been composed by an alternative application of a so-called Cardan Grille, namely by moving a piece of cardboard with holes over a large table of word fragments, and writing down the words that thus appear. This paper caused considerable discussion in the circles of people interested in the Voynich MS text, but it has not found many followers, even until today.

The present paper takes a closer look at the mechanics of this method. Based on this, a more generic method is proposed, which is considerably simpler, both to set up and to execute. It is shown that the unusual word length distribution of the Voynich MS, which is very close to binomial, could be a consequence of the application of such a method. Furthermore, it is argued that this method could not only be used to create meaningless text, but also to encode meaningful text.

A first high-level analysis looks at whether such a method could indeed have been applied to create the Voynich MS text, but this is certainly far from conclusive. The main aim of this paper is to inspire further research of the Voynich MS text into a new direction that has not yet been explored in great detail.

## Introduction

In 2004, Gordon Rugg [1] published a paper in Cryptologia [2], in which he proposed that the Voynich MS text could have been produced using a method that he called the table-and-grille method. It consists of a rather large table of word fragments [3], organised in groups of three columns, and a piece of paper or cardboard with three holes, each hole corresponding to one of the three columns, but at different vertical positions. This piece of paper or cardboard will be referred to in the following as a 'grille'. An example of such a table is given in Figure 1, while some examples of the grilles are given in Figure 2. Figure 3 shows the function of the grille when placed over a table. A later publication by Rugg and Taylor [4] extended this work, and it is also presented on a web site by Hyde and Rugg [5].

---

[1] Dr Gordon Rugg was senior lecturer at Keele University, recently retired
[2] See Rugg (2004)
[3] These word fragments have also been called 'syllables', but I prefer to use the more neutral term
[4] See Rugg and Taylor (2017)
[5] Web page: https://www.hydeandrugg.com/pages/codes/voynich/more-info

Figure 1: a sample table to be used with the table-and-grille method

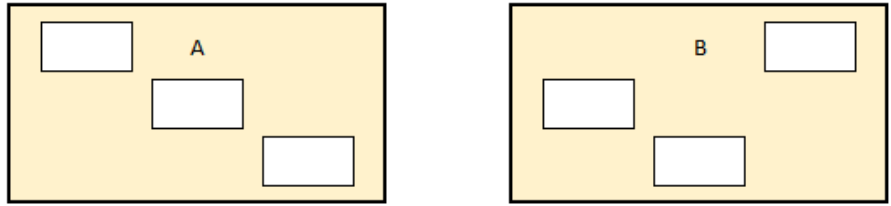

Figure 2: some sample grilles to be used with the table-and-grille method

Figure 3: demonstration of the application of the table-and-grille method

By moving the grille over the table, always aligning it to a group of three columns delimited by the bolder lines, groups of up to three word fragments appear, which then form a Voynich 'word' which should be written down as the next word in the text. Some word fragments may be empty. The principle of composing Voynich words from three word fragments was based on earlier work by the Brazilian mathematician Jorge Stolfi, who analysed the Voynich MS word structure in great detail, and in 1997 proposed [6] that words tend to be composed of three parts which he called prefix, stem and suffix.

Later, in 2000, Stolfi refined his structure [7], which he called 'word grammar', and which was based on three nested levels, which he called core, mantle and crust. The work of Rugg was primarily based on the earlier work of Stolfi.

---

[6] See Stolfi (1997)
[7] See Stolfi (2000a)

The use of grilles, as proposed in the 2004 paper, was based on a cryptographic technique developed by Girolamo Cardano (1501-1576) [8], but is a modification of it. Strictly speaking, its original use was a form of steganography, in which a secret message was hidden in plain view as part of a longer plain text. By placing such a grille over the long text, a new, shorter message appears, which is the secret message that could only be read by someone who had a matching grille. Since this method dates from 1550, Rugg proposed that his alternative technique would have been used by someone after that date, pointing in particular to Edward Kelly, who would have used this method to create a fake manuscript that he would sell for a large sum of money to Emperor Rudolf. This way of producing a text implies that the MS would be a 'hoax', containing meaningless text, even though the perpetrator might have believed that he was divinely inspired in his movements, and create something meaningful. We don't need to be concerned with the historical circumstances of this theory [9], but can concentrate on the mechanism, and see if this is something that could have been applied in practice. The purpose of this paper is also not to address the question of meaningful vs. meaningless. In either case, the text was composed somehow, and this is what we want to look into.

## Could this method have worked in practice?

How would this method have been used in practice? Different possibilities can be envisaged. Either the perpetrator had one very large table, or he had several smaller ones. He is likely to have had several different grilles, and swapped them from time to time. We will see that the frequency of such swapping will have a major impact on understanding how this method could work.

When Rugg first proposed his method, he included some texts generated from sample tables and grilles that, when written using the Voynich MS characters, had the general appearance of the Voynich MS text. That is was not really the same was only clear to people who were very familiar with the Voynich MS text. He never provided an example of a table and a grille that would actually generate the text of one of the pages in the MS. As we will see below, this is actually trivially easy to do, and the table in Figure 1, combined with grille B from Figure 2, can be used to create the text of Voynich MS page f9v.

Let us first assume that the grilles were meant to cover three columns and three rows, and there has to be one hole in each column. This suggests that there would be $3^3 = 27$ different grilles. In reality, the number is smaller, because some cases are duplicates of each other. For this, see Figure 4, where grilles X and Y are effectively the same. The general formula for the number of different configurations is:

$$N = r^c - (r-1)^c$$

where $c$ is the number of columns, which is three and should remain fixed, while $r$ is the number of rows, which is also three, but could be varied. For our example, the answer is $3^3 - 2^3$ = 27 - 8 = 19. If we assume that the perpetrator used all possible grids, then we can compute the size of the table that would be needed.

---

[8] Information according to the Wikipedia entry for "Cardan grille", retrieved 29 March 2021
[9] For more details about this part of the history of the MS, see Zandbergen (2018)

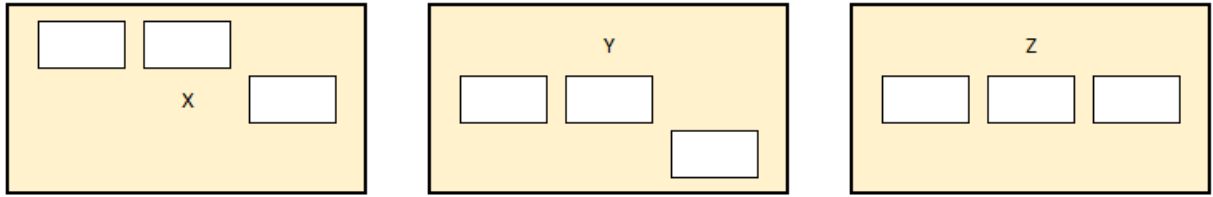

Figure 4: additional sample grilles

The Voynich MS contains approximately 38,000 words in total, while a realistic estimate of the number of different words in the text is 9500 [10] . For the total number of words we may refer to 'word tokens' while for the different words we may refer to 'word types'. The method should be able to generate all word types, so the table (or the combined tables) should have at last 9500 / 19 = 500 entries. 500 entries would require a very large table, but it could easily be split up into three manageable tables, where each table has four sets of columns and 42 rows each: 3 * 4 * 42 = 504. This calculation requires that the table structure is such that every combination using every grid leads to a valid Voynich word, but also that every Voynich word type can only be generated by one grille in one position. Whether that is possible may be left unanswered for the moment. If the same word can be generated by different grilles and/or in different positions, more or larger tables will be required.

We can also re-compute these numbers for grilles with different numbers of rows, for which see Table 1.

Table 1: table length needed for the Voynich MS text as a function of vertical grille size

| Nr. of grille rows | Nr. of different grilles | Table entries |
| --- | --- | --- |
| 2 | 7 | 1358 |
| 3 | 19 | 500 |
| 4 | 37 | 257 |
| 5 | 61 | 156 |

As expected, for larger grilles, the required number of table entries becomes increasingly smaller, and when using grilles with five rows, already a single table of manageable size should in principle be capable of generating all words that appear in the Voynich MS. The constraints on the contents of the table are increasingly important: the word fragments must be increasingly organised in the table, in order to always generate valid words with all possible grilles. Also, there is an increasing 'boundary problem'. The grille must always be completely inside the table, so the number of possible locations to place the grille decreases as the number of rows of the grille increases.

---

[10] These numbers are based on the analysis presented in Zandbergen (2020)

## The effect of changing the grille

Figure 4 includes one grille that seems uninteresting, namely grille Z, where all three holes are aligned horizontally. When using such a grille, composing a table for its use becomes trivial. It just consists of writing out the words that we want to see appear in the MS, arbitrarily split into three parts. Then, moving the grille over the table just means that we select one word after the other.

Obviously, we don't want to use such a grille, but only the ones that have the holes in different vertical positions. However, this is not a significant difference at all!

Using the trivial grille Z with a table of word fragments is fully equivalent with using the better-looking grille Y, by shifting all third columns of word fragments in the table down by one position. It is this simple observation that made it possible to easily generate the table in Figure 1, in combination with grille B in Figure 2, namely by taking all the word types of page f9v in the Voynich MS, breaking them up arbitrarily, and shifting the parts up and down as required for this particular grille.

Clearly, using any single grille for a long time makes this method trivial, while changing the grilles more regularly makes it (apparently) more interesting. Of course, changing from one grille to another is a very easy thing to do, while shifting the columns in the table does not seem to be so easy. However, that is something that can be arranged.

## Changing the table

The table has several sets of three columns next to each other, and thus has a rectangular shape that is convenient to work with. One could imagine, however, that there would just be a single (very long) set of three columns, and even that the beginning and end are connected to each other, forming a circular band with three columns. Doing this does not seem very practical, but at least it removes the (minor) boundary problem that was mentioned before. Effectively, this means that this setup is equivalent, but not completely identical with the use of a large table, where the only difference is the boundary problem.

One can thus picture the table as a (large) wheel. Moving the grille around the table is now equivalent with spinning the wheel. As a next step, one can also imagine that there is not a single wheel, but three separate wheels, one for each column. In this case, changing the grille becomes equivalent with a small shift between the wheels. Thus, the original table-and-grille method is equivalent with the use of three wheels that can be rotated with respect to each other over a very small number of steps.

However, in this scenario, there is no need to limit the relative movement between the wheels to very small steps. They can be extended to the full range of the wheels. In that case, three wheels with only 24 word fragments each, will give rise to 24*24*24 = 13,824 different words, already well in excess of the number of word types in the manuscript. With this change, the table-and-grille method is reduced to a very manageable size. Also, 22 word fragments per table would already be sufficient, as it would generate 10,648 different word types, but we will use 24 in the following, primarily because this brings some practical advantages.

The question remains whether it is possible to subdivide all words in the Voynich MS text consistently into three groups with not more than 24 different components each. From the analysis in Stolfi's 1997 paper, one may have serious doubts about this, as it presents longer tables of word fragments. On the other hand, most entries refer to extremely rare words. We will not try to answer this question now, but only look into it briefly further below, and leave a more detailed analysis of this for future work.

The table-and-grille method of Rugg proposes that this method would have been used to generate a meaningless text. The more recent work of Torsten Timm, while using a very different method [11], equally proposes that words are generated arbitrarily, 'on the fly', and written out as a meaningless stream of words. There is an alternative possibility, namely that either method could have been used to set up a vocabulary of words for an invented language, or as an enumeration system for words, and this vocabulary would have been subsequently used to translate a meaningful text. It may not seem very likely that anyone in the 15th century would have done this, but the same could be said for the arbitrary generation of a string of words using either method. Speculating about the relative likelihoods of these possibilities will not lead to any clear insights, but it is worth to keep in mind that the method using the three wheels, as described above, would be quite capable of generating a meaningful text.

## Can this method explain all properties of the Voynich MS text?

The Voynich MS text has a number of unusual properties, and it is important to see to what extent the proposed three-wheel method would be able to replicate these. One reason for doing this is, that it was already pointed out in the past that the table-and-grille method fails to explain a number of these properties. More in general, most proposed solutions of the Voynich MS that have appeared in recent years tend to ignore these unusual properties, and this is why they either fail to work, or fail to convince.

The table-and-grille method in principle explains, at least to some extent, the most important property of the Voynich MS text: its word structure. Other points that will be looked at are:

- The Currier languages (i.e. the fact that different parts of the text show different properties)
- The preferential placing of some characters, e.g. at the starts of paragraphs, at the top lines of paragraphs and at the ends of lines
- The word length distribution

### *Word structure*

As mentioned before, the table-and-grille method was inspired by an initial version of a word grammar by Jorge Stolfi, whereas he published an improved word grammar later. It will be useful to check if the three-wheel method could be compatible with this later word grammar. Due to the complexity of this word grammar, this analysis can only be made in a qualitative manner. There is certainly scope for significant further analysis in this area.

---

[11] See Timm and Schinner (2019)

As demonstrated by Stolfi [12], the word grammar explains about 96% of all text words, and is based on a three layer model, with a *core* that can be composed of a small set of characters, a surrounding *mantle*, with another limited set of characters, and around that a *crust* with a third set of characters. Figure 5 shows a schematic of this structure, and Table 2 the characters that may appear in each layer.

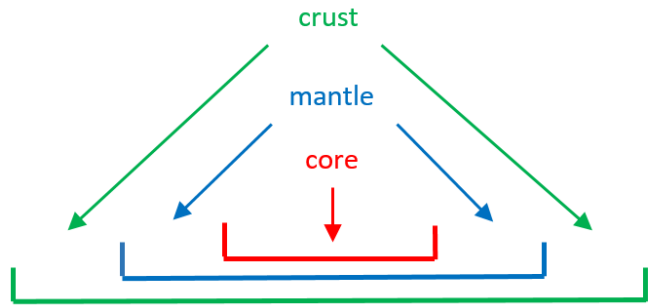

Figure 5: structure of Stolfi's word grammar

Table 2: components of Stolfi's word grammar

| Core   | ተ ቶ ቷ ቸ ቹ ቺ ቻ ቼ |
|--------|------------------|
| Mantle | ᴄᴢ ᴢ̃ ᴄᴄ         |
| Crust  | 8 ɣ ʔ ʕ ʔ ʇ ˌ ʂ ʐ |

There are additional rules for the appearance of the characters ᴀ, ᴏ and ᵩ (which tend to precede any core, mantle or crust character) for the single ᴄ, which tends to follow core or mantle characters, and for the ᛏ, which tends to precede an ᴏ and to appear only at the start of a word. We may consider them to be grouped with these characters, except for the frequent word-final ᵩ, which should be considered a crust character by itself. The grammar says that each level can be empty.

With this word grammar, the three wheels can be considered to consist (in a first iteration) of:

| crust – mantle | mantle - core - mantle | mantle - crust |
|---|---|---|

---

[12] See Stolfi (2000a)

This is clearly a very ambiguous guideline for splitting each Voynich MS word into (up to) three parts, but further analysis might lead to a better general or even specific rule. Just to present one example, the Voynich word [glyph] should be parsed (according to Stolfi's word grammar) as:

[glyph] - [glyph] - [glyph] - [glyph] , which are: core - mantle - crust - crust, and can be assigned to the three wheels in several different ways. One might even consider not to parse [glyph] together with [glyph], but instead to move the [glyph] (or just the [glyph]) to the left wheel. Only a global analysis of all words will be able to demonstrate whether such a split is indeed possible consistently for the majority of Voynich MS words.

### Currier languages

The existence of two different 'languages' in the Voynich MS was first proposed by Currier in 1976 [13]. He hastened to add that these might not really represent different languages in the linguistic sense of the word, but they show some consistent differences, and in most cases, each page of the MS could be associated with either of the two languages, which he called A and B. Later work demonstrated that finer variations in these languages exist, and also some intermediate forms. Most importantly, there is a common basis of words that exists in both languages, and then there is a large group of words that appears only in the B language.

In the case of the original table-and-grille approach, a relatively easy explanation presents itself, namely that different tables were used for the pages in the two languages. However, this would not easily explain that almost all A-language words also tend to appear in the B language.

For the three-wheel method, this problem is a prohibitive one. It is clear that a single set of three wheels, used to arbitrarily generate words, will not lead to two different languages as described above. However, in case this system were used to generate a vocabulary of words, then the two languages might appear as a result of the different vocabulary of the subject matter. This possibility clearly remains open.

### Preferential placing of certain characters

This general heading describes a number of different, yet related features of the text. We will discuss three of these.

The first is that the first word in each paragraph typically starts with a character from a very small group, and this character seems to have been pre-fixed to this word. This character is often written larger than the other characters on the page, and if this character is removed, a regular Voynich word appears, at least in most cases. These words play a special role in Stolfi's word grammar, because they usually don't fit the grammar, when the extra character is included.

---

[13] For Currier's work, see D'Imperio (1976)

It is believed by many that these characters are more of an ornamental nature, and simply mark the start of a new paragraph. This would not be anything mysterious, as this is also done in legible manuscripts of the time. However, both for the table-and-grille method, and the three-wheel method, this would mean that these characters were added separately, and did not come out of the word generation method. This should probably not be seen as a major issue, but speaks against the idea of an arbitrarily-generated text.

The second feature is a much more serious one. This is that the characters  and , and also  and , tend to appear only in top lines of paragraphs. Furthermore, this is not a very hard rule – they do appear elsewhere too, but the predominance in top lines of paragraphs is very strong. This might be just a minor issue, in case they were just alternative forms for ,  (etc.), but this is not the case. This is a prohibitive feature for a large number of proposed solutions to the Voynich MS, and also for the table-and-grille method or the three-wheel method. These characters cannot just appear somewhere in the tables or on the wheels, because then such words could appear anywhere in the text. There would have to be some special rule for the appearance of these characters.

The only 'good' thing in this context is that, indeed, this feature is not yet understood, and a better understanding of the meaning of these characters needs to be found in any case, for all possible explanations of the Voynich text. Once this is found, this problem can be looked at again.

The third feature is similar to the second, but it is less pronounced, and could be easier to explain. This is the character  that is a word-final character that predominantly (but again not always) appears at the ends of lines. In this case, the letter could conceivably be a line final variant form of either  or , but there are some issues with that hypothesis. Again, it would require some special rule for generating the text using any of these methods.

### *Word length distribution*

The last typical feature of the Voynich MS text to be discussed here was discovered by Jorge Stolfi, and this is that the word length distribution for all word types in the text, when using the Eva transliteration alphabet, is almost exactly binomial. His result showed that word lengths range from 1 to 10, with the double peak at 5 and 6 characters. This is shown in Figure 6, which is from Stolfi's web site [14]. This is a striking feature, and not usual for known languages, which tend to have the maximum word frequency (by word length) for relatively shorter words, i.e. the distribution is skewed.

This binomial word length distribution could, however, be explained very well by the three-wheel method described here (and equally by the original table-and-grille method) [15]. In particular, this would be the case if the length distributions of the word fragments of each wheel were also binomial. Let us look at this in more detail.

---

[14] See Stolfi (2000b)
[15] This feature was already known to Rugg and described by him, see Rugg and Taylor 2017

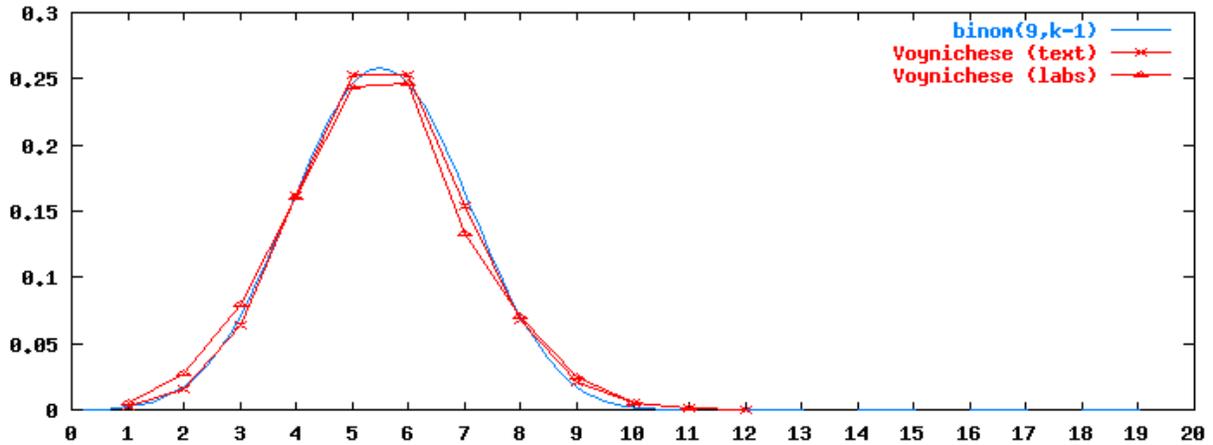

Figure 6: the binomial word length distribution as discovered by Jorge Stolfi

In order to arrive at word lengths ranging from 1-10, the three wheels could have length distributions of 0-3 characters for two of the wheels, and 1-4 character for the fourth. Many other combinations are possible. Table 3 shows this proposed approach in more detail.

Table 3: example of binomial word fragment length distributions for the three wheels

| Wheel | Left | | | | Centre | | | | Right | | | |
|---|---|---|---|---|---|---|---|---|---|---|---|---|
| Fragment length | 0 | 1 | 2 | 3 | 0 | 1 | 2 | 3 | 1 | 2 | 3 | 4 |
| Probability ratio | 1 | 3 | 3 | 1 | 1 | 3 | 3 | 1 | 1 | 3 | 3 | 1 |

The probability ratios for the word fragments in this table (1:3:3:1) correspond to the binomial distribution. For wheels with 24 entries, this translates to the following number of word fragments:

Table 4: word fragment length distribution in terms of number of items

| Wheel | Left | | | | Centre | | | | Right | | | |
|---|---|---|---|---|---|---|---|---|---|---|---|---|
| Fragment length | 0 | 1 | 2 | 3 | 0 | 1 | 2 | 3 | 1 | 2 | 3 | 4 |
| Number | 3 | 9 | 9 | 3 | 3 | 9 | 9 | 3 | 3 | 9 | 9 | 3 |

The word length distribution for all possible combinations of words using such a wheel is exactly binomial. More interestingly, the distributions of the word fragments do not need to be exactly binomial. It is sufficient that they are more or less symmetric and have clearly lower counts for the extremes than in the middle. The combination will still be close to binomial. Table 5 shows two alternative distributions in addition to the binomial case. The total number of word fragments per wheel is always 24.

Table 5: binomial and alternative word fragment length distributions

| Wheel | Left | | | | Centre | | | | Right | | | |
|---|---|---|---|---|---|---|---|---|---|---|---|---|
| **Fragment length** | 0 | 1 | 2 | 3 | 0 | 1 | 2 | 3 | 1 | 2 | 3 | 4 |
| **N (binomial)** | 3 | 9 | 9 | 3 | 3 | 9 | 9 | 3 | 3 | 9 | 9 | 3 |
| **N (alternative 1)** | 2 | 10 | 8 | 4 | 3 | 7 | 10 | 4 | 5 | 7 | 9 | 3 |
| **N (alternative 2)** | 5 | 7 | 7 | 5 | 1 | 11 | 11 | 1 | 4 | 8 | 8 | 4 |

For these three cases, the word length distribution is listed in Table 6 and visualised in Figure 7.

Table 6: word length distributions for the binomial and alternative cases listed in Table 5

| | Binomial | | Alternative 1 | | Alternative 2 | |
|---|---|---|---|---|---|---|
| **Word length** | **Number** | **%** | **Number** | **%** | **Number** | **%** |
| 1 | 27 | 0.20 | 30 | 0.22 | 20 | 0.14 |
| 2 | 243 | 1.76 | 262 | 1.90 | 288 | 2.08 |
| 3 | 972 | 7.03 | 932 | 6.74 | 1092 | 7.90 |
| 4 | 2268 | 16.41 | 2092 | 15.13 | 2284 | 16.52 |
| 5 | 3402 | 24.61 | 3130 | 22.64 | 3228 | 23.35 |
| 6 | 3402 | 24.61 | 3322 | 24.03 | 3228 | 23.35 |
| 7 | 2268 | 16.41 | 2444 | 17.68 | 2284 | 16.52 |
| 8 | 972 | 7.03 | 1204 | 8.71 | 1092 | 7.90 |
| 9 | 243 | 1.76 | 360 | 2.60 | 288 | 2.08 |
| 10 | 27 | 0.20 | 48 | 0.35 | 20 | 0.14 |
| **Total** | **13,824** | **100** | **13,824** | **100** | **13,824** | **100** |

The blue bars labelled "Binom" conform to the word fragment length distribution of Table 4, and is, indeed, exactly binomial. The two alternatives, in orange and grey, are still close to binomial.

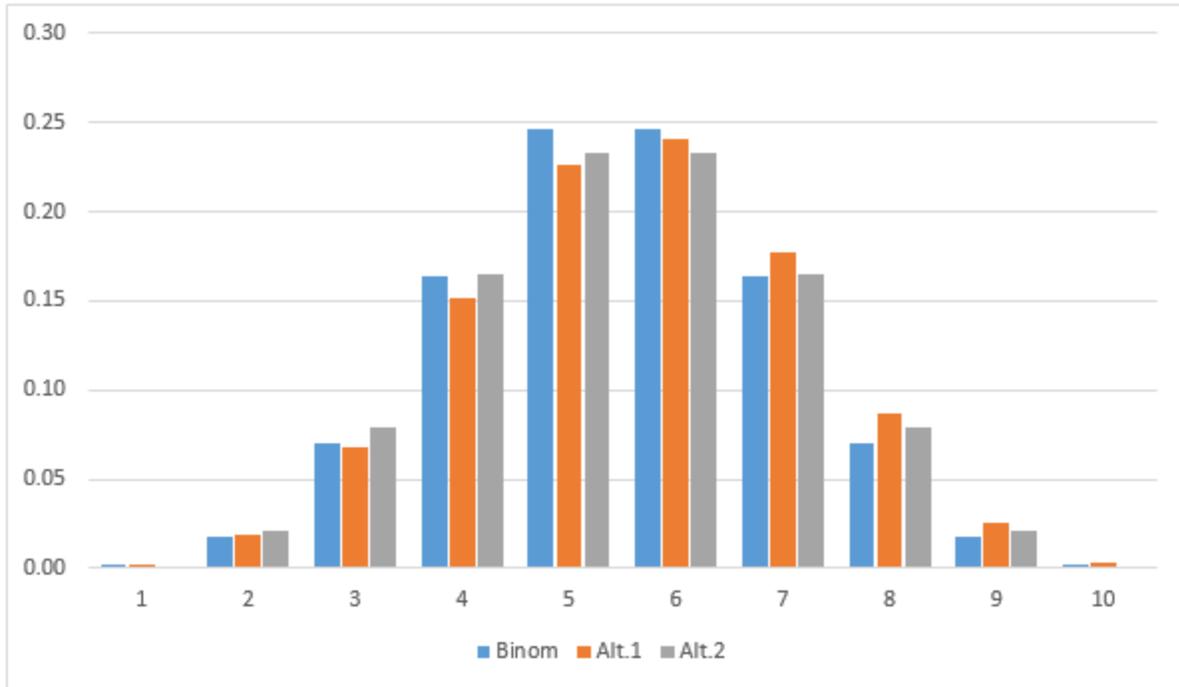

Figure 7: word length distributions for the binomial and alternative cases listed in Table 5

## Some consequences of the approach with three wheels

The mechanism based on three wheels, with 24 word fragments each, would generate 13,824 different words, which will include at least 4000 words that do not actually occur in the Voynich MS text. This is partially a philosophical problem. Any single book written in English will only use a subset of all valid English words. A person who has only this book would not have evidence of all the valid English words that do not occur in it. This is exactly the situation we have with the Voynich MS, provided its text is meaningful. Furthermore, we know for certain that we do not have the complete MS [16], so many words that we are not aware of now must have appeared in it, also in case the text were meaningless. This just means that it is not a major issue if we can define a way to generate all valid Voynich MS words, while this method also generates words that are not attested.

The question remains, how such a method could have been used to encode a meaningful text. A set of wheels as described here generates a list of words that could be considered an enumeration system, simply counting from 1 to some large number N. Hypothetically, someone could have generated a word list or dictionary, to translate words in some existing language to this system. This may seem unlikely from a modern viewpoint, but that is quite subjective. This possibility should not be discarded just yet.

---

[16] At least 14 numbered folios are missing from the MS, but it may well have been incomplete even before the folios were numbered

The choice of three wheels with 24 different entries each also allows other possibilities. The number 24 represents the number of characters in the Greek alphabet, but also Latin (when equating I with J and U with V) and Italian (when not considering k or w). A text in any of these languages could have been split into groups of three characters, where each character would then be mapped to a Voynich word fragment according to the three wheels. This is of course highly speculative.

**Further generalisations**

The approach discussed here can be generalised further, by moving from the concept of three wheels with the same number (24) of word fragments on each of them, to any number of wheels, possibly with different numbers of word fragments. The aspect of having different number of fragments per wheel can be addressed first.

For the case of meaningless text generation, in the original sense of Rugg's table-and-grille method, this would not work. If the number of text fragments per wheel (or per column) would not be the same for all, it would no longer be possible to move the grille around, or to spin the three wheels together. This option only remains in the alternative application, namely setting up a vocabulary of words in an invented or synthetic language. Allowing the possibility to have different numbers of word fragments per wheel, gives more freedom when searching for an algorithm or word structure table that could cover all words in the manuscript.

Table 7: Tiltman's split of words into roots and suffixes

An example of this is a relatively simple approach proposed by John Tiltman in 1967 [17], which is shown in Table 7. This is equivalent with two wheels, where the left wheel has 12 entries and the right wheel has 20 entries. This allows to generate only 240 words, i.e. a very minor subset of all Voynich MS word types.

Going to the other extreme, the maximum number of wheels to generate a binomial word length distribution from 1 to 10 is nine wheels, each with a 50/50 probability of generating zero or one characters, except for one out of nine wheels, that has a 50/50 probability of generating one or two characters. This is visualised in Table 8 below, where the single deviating wheel has been chosen to be wheel nr. 9.

Table 8: a setup with nine wheels

| Wheel | 1 | | 2 | | 3 | | 4 | | 5 | | 6 | | 7 | | 8 | | 9 | |
|---|---|---|---|---|---|---|---|---|---|---|---|---|---|---|---|---|---|---|
| **Fragment length** | 0 | 1 | 0 | 1 | 0 | 1 | 0 | 1 | 0 | 1 | 0 | 1 | 0 | 1 | 0 | 1 | 1 | 2 |
| **Probability** | 1 | 1 | 1 | 1 | 1 | 1 | 1 | 1 | 1 | 1 | 1 | 1 | 1 | 1 | 1 | 1 | 0 | 1 |

From this setup, it is possible to define all possible groupings of wheels. When combining two of these wheels into one, the distribution should become 1:2:1 , when combining three: 1:3:3:1, when combining four: 1:4:6:4:1 etc.

It is clear that Table 3 is just one example of such a grouping, where the wheels 1-3, 4-6 and 7-9 have been combined into one each. The word grammar of Stolfi seems to suggest that an arrangement with five wheels might be one of the more interesting options. Without going into much further analysis, the setup with nine wheels would work if one could find a single ordering of characters in Voynich words that would allow to generate all words, simply by omitting some of these characters. It is not clear that this is possible, but it is clear at the same time that there are more than nine different characters in the Voynich alphabet, so some characters would need to share the same wheel. There are clear candidates for this, because it is a well-known property of the Voynich MS text that certain characters can be replaced by others, and still generate a valid word. Examples are: ⊘/⊘ , ⟨⟨/⟨⟨ , ?/?/? , ⟨⟨ .

This arrangement with nine wheels would explain one more intriguing property of the Voynich MS text, which has been explored extensively in the work of Timm, namely the fact that the corpus of words in the Voynich MS forms a network of words connected by small edit distances [18]. Without going into great detail, this means that almost all words in the MS can be converted into other valid words just by deleting, changing or adding one character. The same could be achieved by arrangements with fewer wheels, but with some constraints on the possible contents of these wheels.

---

[17] See Tiltman (1967)
[18] See Timm and Schinner (2019)

In case the wheels are used to set up an enumeration system, it is not even needed to actually create the wheels and turn them. The combinations can be written down without this. This may be illustrated by Table 9, which shows a set of wheels that would generate all Roman numerals up to 4999 [19].

Table 9: A set of seven wheels that would generate Roman numerals

| - | - | - | - | - | - | - |
|---|---|---|---|---|---|---|
| M | D | C | L | X | V | I |
| MM |  | CC |  | XX |  | II |
| MMM |  | CCC |  | XXX |  | III |
| MMMM |  | CCCC |  | XXXX |  | IIII |

## Concluding remarks

In this paper we have demonstrated that the table-and-grille method proposed by Rugg in 2004 can be seen as a specific case of a more general approach with three wheels with word fragments. While the tables would end up being very large, the wheels could potentially allow for a much more compact solution. For example, the table in Figure 1 is about as large as the full text of the page that it generates. However, it is still to be explored if such a more compact solution to generate the full MS text can be found.

We have then addressed, only at a high level, the important question whether such a method could have been used in practice to generate the Voynich MS text. In this, we have not tried to answer the question whether the text is meaningful or not. This has been left open, an approach that can be justified by the fact that, also in the case that the text is meaningless, there must have been a method of generating it, given all its regularities.

The array of different possible arrangements of wheels allows to explain some of the most important features of the text of the Voynich MS: the word structure, the binomial word length distribution, and the network of words with small edit distances. One prohibitive problem has been identified, namely the special role of the characters ⚨ and ⚩. This is a serious problem, that will require a better understanding of the role of these characters. This is true for all possible solutions of the Voynich MS.

Another problem that needs to be addressed is that the word grammar of Stolfi cannot explain about 4% of all words in the MS. It would be only too easy to assume that these are all either scribal errors or transliteration errors. The former are very likely to exist and the latter certainly exist, but this would not be very convincing. Stolfi identifies several categories of non-matching words, and one thing that emerges is that the word spaces in the MS, both as we see them and as they have been transcribed, are very uncertain. Some exceptional words look like fragments of complete words, or like concatenations of words. We would recommend that the issue of uncertain word spaces is analysed in more detail, based on Stolfi's proposed word grammar, with some allowance for doubting word spaces as observed until now.

---

[19] That is, without using the subtractive notation, which allows to write IV for 4 and IX for 9. This additive-only definition of Roman numerals has also been used occasionally in the past.

As a closing remark, this paper may hopefully provide some new avenues for further research.

## Acknowledgments

The Voynich MS is preserved at the Beinecke Rare Book and MS Library of Yale University, New Haven (CT). I wish to express my gratitude to Gordon Rugg for his careful review of this document, and his constructive suggestions. I am grateful to Jorge Stolfi for his permission to use the illustrations presented at his web site.